\documentclass[12pt,a4paper]{article}
\usepackage[dvips]{epsfig}
\usepackage{graphicx}
\usepackage{multicol}

\title{Energy dependence of\\ 
transverse mass spectra of $\Lambda$ hyperons\\ 
produced in p+p 
and p+$\overline{\rm{p}}$ interactions.\\ A compilation.}
\author{Frederick Kramer$^{a}$, Claudia Strabel$^{a}$ \\
and Marek Ga\'{z}dzicki$^{a,b}$}
\date{\today}

\begin{document}
\maketitle

%-----------------------------------------------------------------------------------------

\noindent
\begin{minipage}[t]{12.5cm}

$^a$ Institut f\"ur  Kernphysik, Universit\"at  Frankfurt,
Germany\\
$^b$ \'Swi\c{e}tokrzyska Academy, Kielce, Poland \\
\end{minipage}

\vspace*{0.4cm}

%------------------------------------------------------------------------------------------

\begin{abstract}
The results on transverse mass spectra of
$\Lambda$ hyperons
produced in all inelastic p+p and p+$\overline{\rm{p}}$
interactions in the energy range $\sqrt{s_{NN}}=3.6-1800$~GeV
are compiled and analyzed.
The energy dependence of the mean transverse mass and the inverse slope parameter of the spectra is presented and discussed.
These results should be used as a reference in the study of $\Lambda$ hyperon production in nuclear collisions.

\end{abstract}

\section{Introduction}

Recently observed anomalies in the energy dependence
of hadron production in central Pb+Pb
collisions
\cite{Afanasiev:2002mx}
suggest that the onset of deconfinement
may be located at the low CERN SPS energies
$\sqrt{s_{NN}} \approx 7.5$ GeV 
\cite{Gazdzicki:1998vd,Gorenstein:2003cu}.
In the identification of the effects possibly related to the onset
of deconfinement in heavy ion collisions a
comparison with the
corresponding results obtained in nucleon-nucleon
interactions plays a special role.
In case of the study of energy dependence of pion and kaon
multiplicities this comparison was done based on the existing compilations of the data on
all inelastic proton-proton (p+p),
proton-neutron and proton-antiproton (p+$\overline{\rm{p}}$) interactions
\cite{Gazdzicki:zs}.

A recently found
\cite{Gorenstein:2003cu}
 anomaly in the energy
dependence of the shape of transverse
mass ($m_T$) spectra of hadrons
produced in central Pb+Pb collisions raised significant interest
[5-8].
It motivates an effort to compile the corresponding results in p+p($\overline{\rm{p}}$) interactions.
The first compilation of the transverse mass spectra of kaons was already published
\cite{Kliemant:2003sa}.\\
In this paper the compilation of the $m_T$ spectra of $\Lambda$ hyperons is presented and discussed.
The paper is organized as follows.
In section~2 the existing data are reviewed and analyzed.
The energy dependence of the spectra is presented
and discussed
in section~3. The paper is closed by the summary,
section~4.\\

\section{A compilation of p+p($\overline{\textbf{p}}$) data}

We start from review of the data on $\Lambda$ hyperon transverse
momentum ($p_T$) and transverse mass spectra in all inelastic
p+p($\overline{\rm{p}}$) interactions.

In most cases the original experimental papers present $p_T$ spectra in the form
$\frac{d^{3}n}{dp^2_Tdy}$ or $\frac{Ed^3n}{d\vec{p}}$.
From these results the transverse mass
$m_T$
($m_T=\sqrt{p^2_T+m^2_0}$, where $m_0$ is the particle rest mass)
spectra $\frac{d^{2}n}{m_{T}dm_{T}dy}$ can be easily obtained
($\frac{d^{2}n}{m_{T}dm_{T}dy}\sim\frac{d^{3}n}{dp^2_Tdy}\sim\frac{Ed^3n}{d\vec{p}}$).

Our compilation and analysis is limited to the low $p_T$ region,
$p_T\leq1.4$~GeV/c.
First, it is because in this region the $p_T$ spectra of $\Lambda$ hyperons were measured by many experiments
and therefore a systematic study is possible.
Second, this is the region where the shape of the spectra in A+A collisions is determined by the hydrodynamical expansion
of the matter and thus it is sensitive to the change of the equation of state due to the onset of deconfinement.
In this region an exponential parameterization
of the p+p($\overline{\rm{p}}$) data:
\begin{equation}
\label{eq1}
\\
\frac{dn}{m_{T}dm_{T}}=C\cdot e^{-m_{T}/T}
\\
\end{equation}
is approximately valid\footnote{
At high $m_T$ the spectra obey a power law behavior,
$\frac{dn}{m_{T}dm_{T}} \sim m_{T}^{-P}$,
which is interpreted as due to
hard (parton) scattering.}.
In Eq.~\ref{eq1} the inverse slope parameter $T$
and the normalization parameter $C$
are free parameters and their values are extracted from
the least square fits to the experimental spectra.

Most of the compiled $p_T$ spectra are measured either close to mid-rapidity (y$\approx0$)
or they are integrated
over forward or backward hemispheres.

Transverse momentum spectra of $\Lambda$ hyperons produced in
all inelastic p+p interactions
are measured in fixed target
[10-23]
mostly bubble chamber experiments,
at energies below $\sqrt{s_{NN}}\approx30$~GeV.
At higher energies, $\sqrt{s_{NN}}= 44-200$~GeV, the data on the spectra are taken
in collider experiments at ISR
[24, 25]
and RHIC
[26]
. The measurements for
p+$\overline{\rm{p}}$ interactions are performed at the Sp$\overline{\rm{p}}$S and Fermilab colliders
\cite{Ansorge:1989ba,Banerjee:1988ps}
at $\sqrt{s_{NN}}= 200-1800$~GeV.
All results come from the analysis of charged decays of
$\Lambda$ hyperons,
$\Lambda$~$\rightarrow p+\pi^{-}$.

The summary of the data on $m_{T}$ spectra of $\Lambda$
hyperons is given in Tables 1 and 2, where
$\sqrt{s_{NN}}$, the 
$p_{T}$ range selected for the analysis, the longitudinal acceptance in which
the measurement was done and the 
reference to the original experimental papers are given. 
The $m_T$ spectra are plotted as a
function of ($m_T-m_0$) in Figs.~1-3.
The normalization of the spectra is arbitrary. They are ordered from
bottom to top according to increasing energy.

The spectra displayed in Figs.~1-3 are fitted by an exponential function, Eq.~\ref{eq1},
in the whole $m_T$ range (($m_T-m_0)\leq0.7$~GeV/c$^2$).
The inverse slope parameter $T$ and $\chi^{2}/NDF$ resulting from the fits are given in Tables~1 and 2.
The corresponding functions are plotted in Figs.~1-3 by solid lines.
It is seen that the used parametrization (Eq.~\ref{eq1}), 
reasonably well describes $\Lambda$ spectra in the 
studied $m_T$ range at all energies
($\sqrt{s_{NN}}=3.6-1800$~GeV), both
for p+p and p+$\overline{\rm{p}}$ interactions.\\
The mean $m_T$ in the range $0<(m_T-m_0)<0.7$ GeV/c$^{2}$ was calculated using measured data.
If necessary, the results were corrected for the unmeasured tails of the distributions by use of the exponential function.
The resulting values of ($<m_T>-m_0$) are also given in Tables 1 and 2.

\section{The energy dependence}

The  energy dependence of
the $T$ and ($<m_T>-m_0$) parameters calculated in the $m_T$ interval
($m_T-m_0)\leq0.7$~GeV/c$^2$
to the $m_T$ spectra of
$\Lambda$ hyperons produced
in all inelastic p+p($\overline{\rm{p}}$) interactions
are shown in Figs.~4 and 5.
The results were parametrized by an expression:
\begin{equation}
\label{eq2}
\\
y=a+b\cdot\ln{(\sqrt{s_{NN}}/(1GeV))},
\\
\end{equation}
where $y=T$ or ($<m_T>-m_0$), $a$ and $b$ are fit parameters
and $\sqrt{s_{NN}}$ is given in units of GeV.
The best fit to the inverse slope parameter data presented in Fig.~4 yields $a=(53.6\pm3.7)$~MeV, $b=(25.4\pm2.0)$~MeV
and $\chi^{2}/NDF=58/23$.
The parameters fitted to the mean transverse mass data are $a=(60.7\pm4.3)$~MeV, $b=(26.7\pm2.2)$~MeV
and $\chi^{2}/NDF=31/23$.

In both cases the observed deviations of points from the parameterizations are mostly consistent with ones
expected from the experimental errors.

The compiled results indicate that $T$ and ($<m_T>-m_0$) increase monotonically with the collision energy from
$T\approx90$~MeV and ($<m_T>-m_0)\approx100$~MeV at $\sqrt{s_{NN}}=3.6$~GeV to $T\approx280$~MeV and ($<m_T>-m_0)\approx350$~MeV at $\sqrt{s_{NN}}=1800$~GeV.
No significant differences are seen between data for p+p and p+$\overline{\rm{p}}$
interactions and between mid-rapidity and integrated data.
Note that in general one expects flatter spectra for mid-rapidity results, this trend is suggested
by several low energy points.

Data of higher statistical precision are needed to draw firm conclusions
on the details of energy dependence of the shape of $m_T$ spectra in p+p($\overline{\rm{p}}$) interactions.
The measurements of the $m_T$ spectra in nucleus-nucleus collisions are still
sparse, but the new results at SPS and RHIC energies are expected soon. The measured
values of the $T$ parameter in central Pb+Pb (Au+Au) collisions are in the range $T=200-350$~MeV at $\sqrt{s_{NN}}=5-200$~GeV
[29-34].
They are significantly larger than the corresponding p+p($\overline{\rm{p}}$) results.

\section{Summary}

We compiled and analyzed data on $m_{T}$ spectra  of 
$\Lambda$ hyperons produced in all inelastic p+p and p+$\overline{\rm{p}}$ interactions.
The spectra can be reasonably well described by a simple exponential parametrization 
$\frac{dn}{m_{T}dm_{T}}=C\cdot e^{-m_{T}/T}$
in the whole analyzed $m_T$ interval ($m_{T}-m_{0}\leq$ 0.7 $GeV/c^{2}$).
We do not observe any significant difference between data from p+p and p+$\overline{\rm{p}}$ interactions
 and results obtained in different rapidity or $x_{F}$ acceptances.
The $T$ and ($<m_T>-m_0$) parameters
increase monotonically with $\sqrt{s_{NN}}$ from $T\approx90$~MeV and ($<m_T>-m_0)\approx100$~MeV
at $\sqrt{s_{NN}}=3.6$~GeV to $T\approx280$~MeV and ($<m_T>-m_0)\approx350$~MeV at $\sqrt{s_{NN}}\approx1800$~GeV.
This dependence can be parametrized as $T=(53.6\pm3.7)$~MeV~+~$(25.4\pm2.0)$~MeV~$\cdot\ln(\sqrt{s_{NN}}/1$~GeV) and  ($<m_T>-m_0)=(60.7\pm4.3)$~MeV~+~$(26.7\pm2.2)$~MeV~$\cdot\ln(\sqrt{s_{NN}}/1$~GeV), where $\sqrt{s_{NN}}$ is given in units of GeV.

These results should serve as a reference in the study of $\Lambda$ hyperon production in nuclear collisions.

\vspace{0.3cm}

\noindent {\bf Acknowledgments}\\
We thank Christoph Blume for discussion and help and our
colleagues at IKF, especially Benjamin Lungwitz and Michael Kliemant.

This work was supported by the Virtual Institute VI-SIM (VI-146) of Helmholtz Gemeinschaft, Germany.

%-----------------------------------------------------------------------------------------

%\pagebreak

%---------------------------------------------------------------------------------------------

\pagebreak

\begin{table}[htp]
\caption{Summary of the data on the $p_{T}$ spectra of $\Lambda$ hyperons produced
in p+p interactions. The collision energy $\sqrt{s_{NN}}$, the $p_{T}$-range used for this
analysis, the rapidity (in center of mass system) or $x_{F}$ range
in which the $p_T$ spectrum was measured
and the references to the original papers are given.
The fitted inverse slope parameter $T$, the corresponding $\chi^{2}/NDF$ and ($<m_{T}>-m_{0}$) are
also presented.}
%~\\[0.1cm]
\hspace{1cm}

\small
\begin{tabular}{|c|c|c|c|c|c|c|}
  \hline
  &&&&&&\\
  $\sqrt{s}$&p$_{t}$-range&longitudinal&&&$<m_{T}>$-m$_{0}$&\\
  (GeV)&(GeV/c)&acceptance&$T$(MeV)&$\chi^{2}/NDF$&(MeV)&ref.\\
  &&&&&&\\
  \hline
  3.63&0-0.94&-0.25$<$ $x_{F}$ $<$0.25&91.8 $\pm$ 4.5&8.1/10&99.1 $\pm$ 5.9&\cite{Eisener:1977mx}\\
  3.78&0-0.97&0$<y<$1.5&81.1 $\pm$ 2.0&10.5/8&89.3 $\pm$ 3.1&\cite{Blobel:1973jc}\\
  4.93&0-1.00& $x_{F}$=0&123.6 $\pm$ 14.1&0.9/7&127.4 $\pm$ 14.9&\cite{Blobel:1978yj}\\
  4.93&0-1.02&0$<y<$1.5&99.5 $\pm$ 2.3&2.4/6&109.1 $\pm$ 3.6&\cite{Blobel:1973jc}\\
  5.01&0-1.15&-1.6$<y<$0&102.8 $\pm$ 7.1&6.1/6&111.4 $\pm$ 10.0&\cite{Jaeger:1974in}\\
  6.84&0-1.03&$x_{F}$=0&146.4 $\pm$ 11.6&5.1/7&155.9 $\pm$ 13.4&\cite{Blobel:1978yj}\\
  6.84&0-1.02&0$<y<$1.9&103.0 $\pm$ 3.6&4.2/6&112.6 $\pm$ 4.7&\cite{Blobel:1973jc}\\
  11.56&0-0.81&0$<y<$2.4&106.2 $\pm$ 15.0&1.8/3&118.8 $\pm$ 21.7&\cite{Blumenfeld:1973vm}\\
  11.56&0-0.92&-2.5$<y<$0&114.0 $\pm$ 17.3&2.1/9&124.2 $\pm$ 16.4&\cite{Ammosov:1975bt}\\
  13.90&0-0.94&-2.2$<y<$0&186.5 $\pm$ 43.6&0.6/4&205.5 $\pm$ 34.7&\cite{Chapman:1973fn}\\
  16.66&0-1.12&0$<y<$0.6&134.1 $\pm$ 18.0&2.5/7&152.4 $\pm$ 30.4&\cite{Brick:1980vj}\\
  19.66&0-1.02&0$<y<$3.0&118.9 $\pm$ 28.7&0.5/2&118.5 $\pm$ 32.8&\cite{Blobel:1973jc}\\
  19.66&0-1.11&-3.0$<y<$0&137.4 $\pm$ 24.1&6.8/5&115.1 $\pm$ 22.8&\cite{Jaeger:1974pk}\\
  23.76&0-1.13&0$<y<$3.0&119.6 $\pm$ 10.1&2.1/5&129.6 $\pm$ 14.2&\cite{Lopinto:1980ct}\\
  23.76&0-1.00&-3.1$<y<$0&171.1 $\pm$ 42.5&3.6/5&185.9 $\pm$ 31.8&\cite{Sheng:1974zn}\\
  26.03&0-1.32&-3.2$<y<$0&123.5 $\pm$ 8.7&5.5/8&137.1 $\pm$ 11.1&\cite{Asai:1984dv}\\
  27.60&0-1.28&-3.4$<y<$0&125.8 $\pm$ 10.9&4.1/3&141.4 $\pm$ 20.3&\cite{Kichimi:1977rj}\\
  27.60&0-1.34&0$<y<$3.0&127.0 $\pm$ 12.2&9.0/8&157.2 $\pm$ 16.5&\cite{Kichimi:1979te}\\
  44.00&0.71-1.41&$y$$\approx$0&130.5 $\pm$ 16.8&6.4/2&206.3 $\pm$ 35.5&\cite{Busser:1975tj}\\
  63.00&0.71-1.21&$y$=0&132.2 $\pm$ 20.7&0.6/1&176.4 $\pm$ 22.6&\cite{Drijard:1981wg}\\
  200.00&0.44-1.40&-0.5$<y<$0.5&235.9 $\pm$ 19.9&0.3/4&244.6 $\pm$ 17.7&\cite{Adams:2004ec}\\
  \hline
\end{tabular}
%\\[1cm]
\end{table}
\normalsize

\begin{table}[htp]
\caption{Summary of the data on the $p_{T}$ spectra of $\Lambda$ hyperons produced
in p+$\overline{\rm{p}}$ interactions. For details see the caption of Table 1.}
%~\\[0.1cm]
\hspace{1cm}

\small
\begin{tabular}{|c|c|c|c|c|c|c|}
  \hline
  &&&&&&\\
  $\sqrt{s}$&p$_{t}$-range&longitudinal&&&$<m_{T}>$-m$_{0}$&\\
  (GeV)&(GeV/c)&acceptance&$T$(MeV)&$\chi^{2}/NDF$&(MeV)&ref.\\
  &&&&&&\\
  \hline
  200&0.64-1.34&0$<y<$2&371.5 $\pm$ 157.8&1.2/2&404.2 $\pm$ 78.5&\cite{Ansorge:1989ba}\\
  546&0.32-1.37&0$<y<$2&235.3 $\pm$ 52.5&1.6/3&229.7 $\pm$ 45.7&\cite{Ansorge:1989ba}\\
  900&0.64-1.43&0$<y<$2&259.5 $\pm$ 53.9&2.2/2&271.9 $\pm$ 41.8&\cite{Ansorge:1989ba}\\
  1800&0.47-1.24&-0.36$<\eta<$1.0&277.0 $\pm$ 103.8&0.0/1&345.8 $\pm$ 108.0&\cite{Banerjee:1988ps}\\
  \hline

\end{tabular}
\normalsize
%\\[1cm]
\end{table}

\newpage

%--------------------------------------------------------------------------------------------------------------
\begin{figure}[p]
  \label{mtspec1}
  \includegraphics[width=11.5cm]{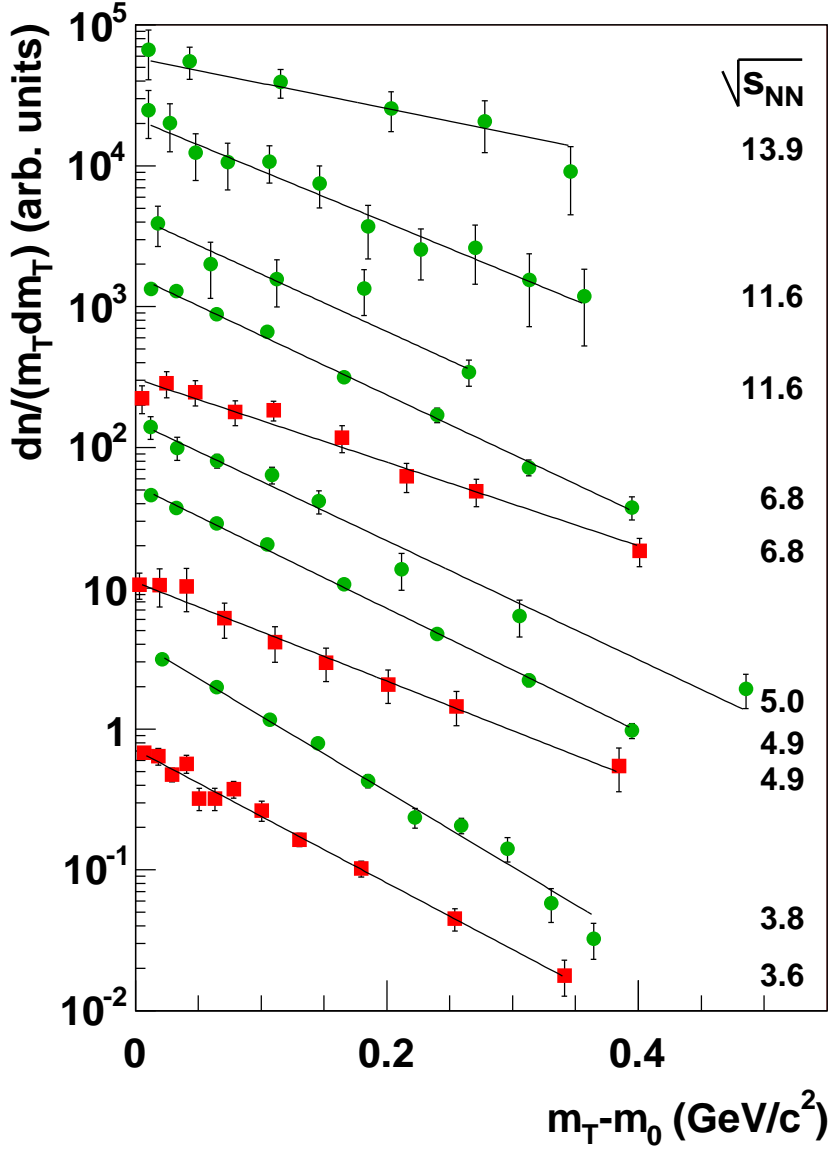}
  \caption{(Color online) Transverse mass spectra of $\Lambda$ hyperons produced in p+p interactions at $\sqrt{s_{NN}}=3.6-13.9$~GeV.
  The spectra are measured either at mid-rapidity (squares) or integrated over semi-hemisphere (circles).
  The normalization of the spectra is arbitrary and the numbers next to the spectra give the c. m. collision energy in GeV.
  They are ordered from bottom to top according to rising energy.
  The fits of the exponential function (Eq.~\ref{eq1}) are indicated by the solid lines.}
\end{figure}

\begin{figure}[p]
  \label{mtspec2}
  \includegraphics[width=11.5cm]{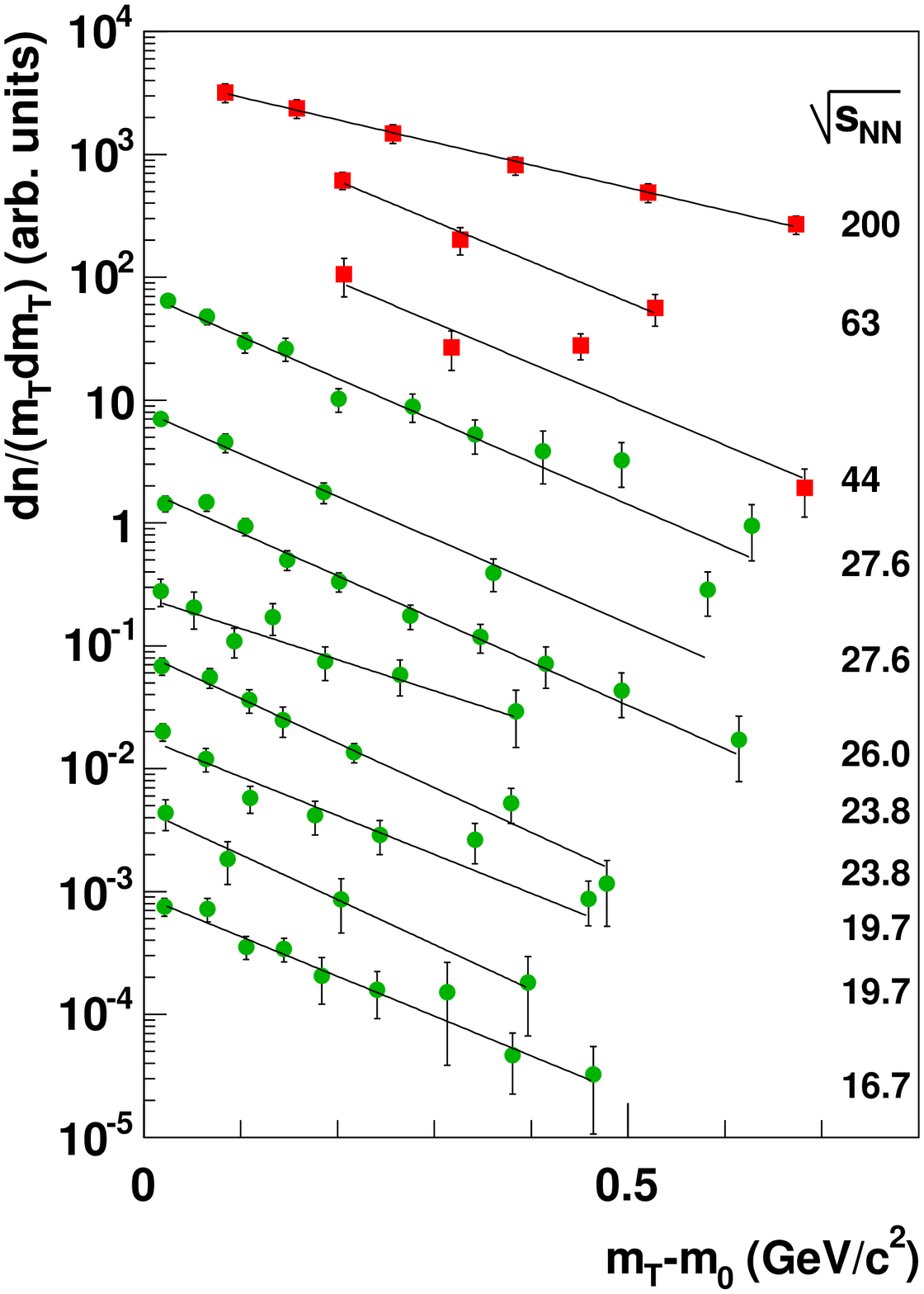}
  \caption{(Color online) Transverse mass spectra of $\Lambda$ hyperons produced in p+p interactions at $\sqrt{s_{NN}}=16.7-200$~GeV.
  For details see the caption of Fig.~1.}
\end{figure}

\begin{figure}[p]
  \label{mtspec3}
  \includegraphics[width=11.5cm]{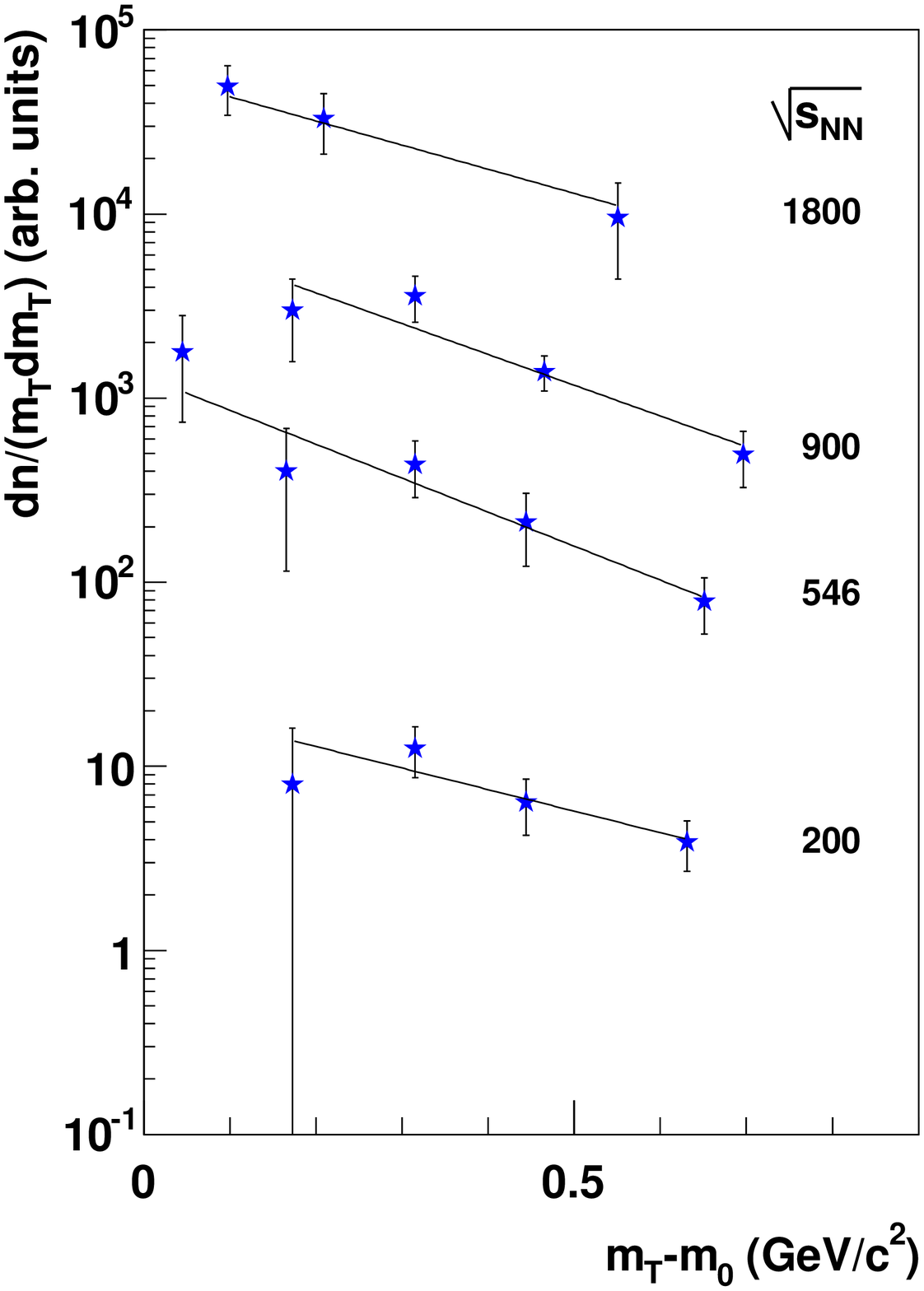}
  \caption{(Color online) Transverse mass spectra of $\Lambda$ hyperons produced in p+$\overline{\rm{p}}$ interactions at $\sqrt{s_{NN}}=200-1800$~GeV.
   For details see the caption of Fig.~1.}
\end{figure}

\begin{figure}[p]
  \label{ppenergy}
  \includegraphics[width=14cm]{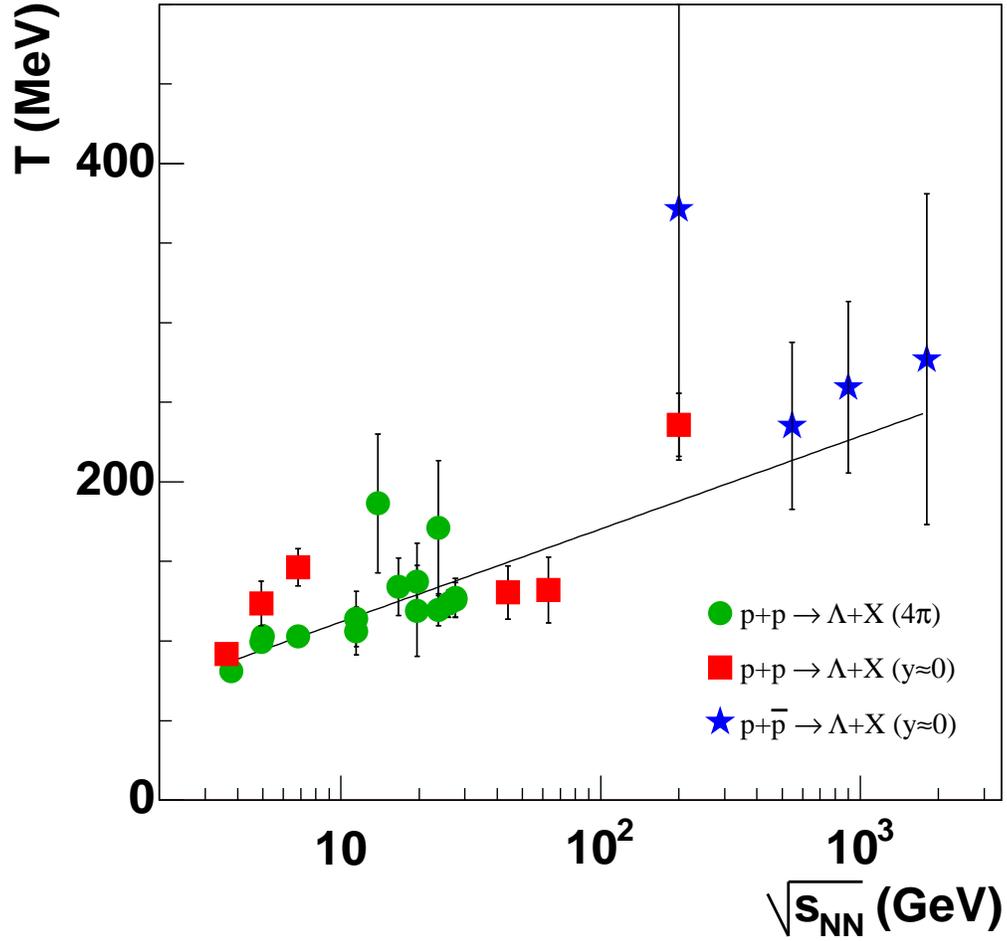}
  \caption{(Color online) Energy dependence of
  the inverse slope parameter $T$ of transverse mass
  spectra of $\Lambda$ hyperons produced
  in p+p and p+$\overline{\rm{p}}$ interactions.
  The $T$ parameter was determined by fitting the spectra
  (Eq. 1) in the whole analyzed $m_T$ interval,
  $(m_T -m_0) \leq 0.7$ GeV/c$^2$.
  The logarithmic parameterization is indicated by the solid line.}
\end{figure}

\begin{figure}[p]
  \label{meanmt}
  \includegraphics[width=14cm]{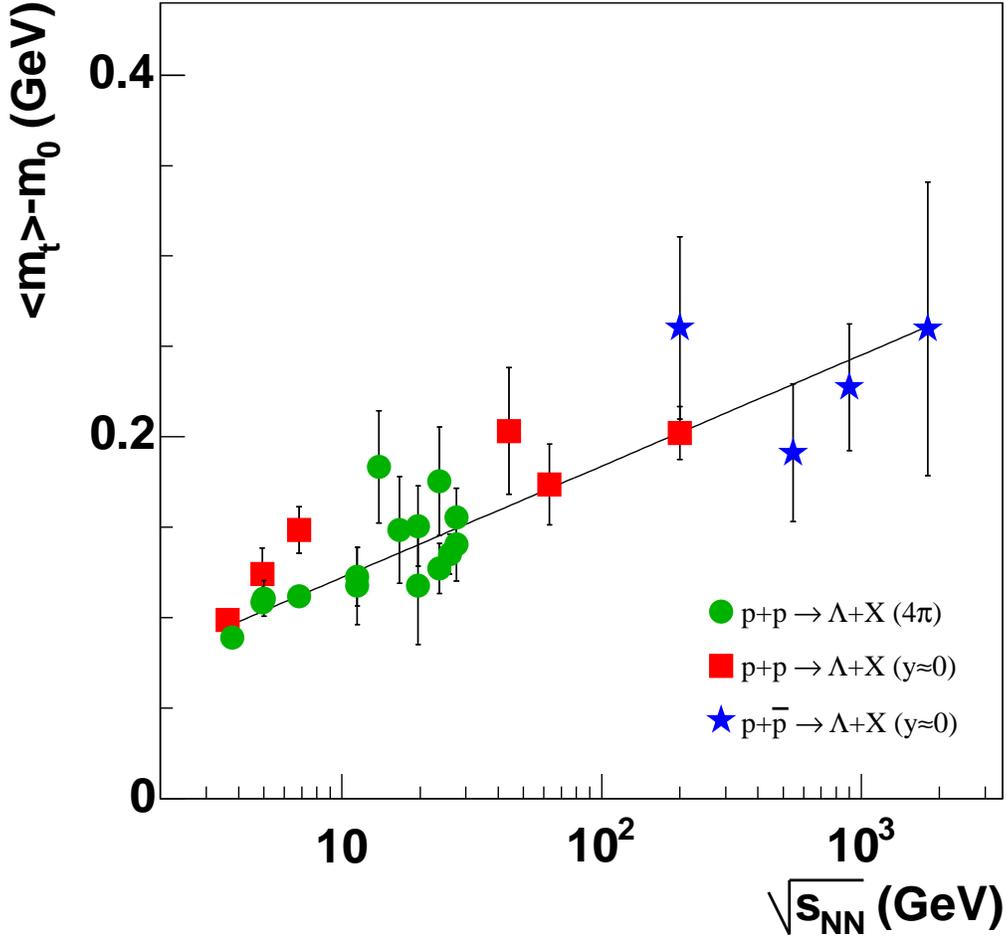}
  \caption{(Color online) Energy dependence of the mean transverse mass of $\Lambda$ hyperons produced
  in p+p and p+$\overline{\rm{p}}$ interactions. The ($<m_{T}>-m_{0}$) was calculated in the range $(m_T -m_0) \leq 0.7$ GeV/c$^2$.
  The logarithmic parameterization is indicated by the solid line.}
\end{figure}

\end{document}